\documentclass[%
 reprint,
superscriptaddress,
nofootinbib,
nobibnotes,
 amsmath,amssymb,
 aps,
]{revtex4-1}

\usepackage{graphicx}
\usepackage{dcolumn}
\usepackage{bm}
\usepackage{multirow}
\usepackage{color}

\begin{document}

\preprint{APS/123-QED}

\title{Search for three body pion decays ${\pi}^+{\to}l^+{\nu}X$}

\author{A.~Aguilar-Arevalo}
\affiliation{Instituto de Ciencias Nucleares, Universidad Nacional Aut\'onoma de M\'exico, CDMX 04510, M\'exico}

\author{M.~Aoki}
\affiliation{Department of Physics, Graduate School of Science, Osaka University, Toyonaka, Osaka, 560-0043, Japan}

\author{M.~Blecher}
\affiliation{Virginia Tech., Blacksburg, Virginia 24061, USA}

\author{D.I.~Britton}
\affiliation{SUPA - School of Physics and Astronomy, University of Glasgow, Glasgow, G12-8QQ, United Kingdom}

\author{D.~vom~Bruch}
\thanks{Present address: LPNHE, Sorbonne Universit\'e, Universit\'e Paris Diderot, CNRS/IN2P3, Paris, France.}
\affiliation{Department of Physics and Astronomy, University of British Columbia, Vancouver, British Columbia V6T 1Z1, Canada}

\author{D.A.~Bryman}
\affiliation{Department of Physics and Astronomy, University of British Columbia, Vancouver, British Columbia V6T 1Z1, Canada}
\affiliation{TRIUMF, 4004 Wesbrook Mall, Vancouver, British Columbia V6T 2A3, Canada}

\author{S.~Chen}
\affiliation{Department of Engineering Physics, Tsinghua University, Beijing, 100084, China}

\author{J.~Comfort}
\affiliation{Physics Department, Arizona State University, Tempe, AZ 85287, USA}

\author{S.~Cuen-Rochin}
\affiliation{TRIUMF, 4004 Wesbrook Mall, Vancouver, British Columbia V6T 2A3, Canada}
\affiliation{Universidad Aut\'onoma de Sinaloa, Culiac\'an, M\'exico}

\author{L.~Doria}
\affiliation{TRIUMF, 4004 Wesbrook Mall, Vancouver, British Columbia V6T 2A3, Canada}
\affiliation{PRISMA$^+$ Cluster of Excellence and Institut f\"ur Kernphysik, Johannes Gutenberg-Universit\"at Mainz, Johann-Joachim-Becher-Weg 45, D 55128 Mainz, Germany}

\author{P.~Gumplinger}
\affiliation{TRIUMF, 4004 Wesbrook Mall, Vancouver, British Columbia V6T 2A3, Canada}

\author{A.~Hussein}
\affiliation{TRIUMF, 4004 Wesbrook Mall, Vancouver, British Columbia V6T 2A3, Canada}
\affiliation{University of Northern British Columbia, Prince George, British Columbia V2N 4Z9, Canada}

\author{Y.~Igarashi}
\affiliation{KEK, 1-1 Oho, Tsukuba-shi, Ibaraki, 300-3256, Japan}

\author{S.~Ito}
\thanks{Corresponding author (s-ito@okayama-u.ac.jp).\\Present address: Faculty of Science, Okayama University, Okayama, 700-8530, Japan.}
\affiliation{Physics Department, Osaka University, Toyonaka, Osaka, 560-0043, Japan}

\author{S.~Kettell}
\affiliation{Brookhaven National Laboratory, Upton, NY, 11973-5000, USA}

\author{L.~Kurchaninov}
\affiliation{TRIUMF, 4004 Wesbrook Mall, Vancouver, British Columbia V6T 2A3, Canada}

\author{L.S.~Littenberg}
\affiliation{Brookhaven National Laboratory, Upton, NY, 11973-5000, USA}

\author{C.~Malbrunot}
\thanks{Present address: Experimental Physics Department, CERN, Gen\`eve 23, CH-1211, Switzerland.}
\affiliation{Department of Physics and Astronomy, University of British Columbia, Vancouver, British Columbia V6T 1Z1, Canada}

\author{R.E.~Mischke}
\affiliation{TRIUMF, 4004 Wesbrook Mall, Vancouver, British Columbia V6T 2A3, Canada}

\author{T.~Numao}
\affiliation{TRIUMF, 4004 Wesbrook Mall, Vancouver, British Columbia V6T 2A3, Canada}

\author{D.~Protopopescu}
\affiliation{SUPA - School of Physics and Astronomy, University of Glasgow, Glasgow, G12-8QQ, United Kingdom}

\author{A.~Sher}
\affiliation{TRIUMF, 4004 Wesbrook Mall, Vancouver, British Columbia V6T 2A3, Canada}

\author{T.~Sullivan}
\thanks{Present address: Department of Physics, University of Victoria, Victoria BC V8P 5C2, Canada.}
\affiliation{Department of Physics and Astronomy, University of British Columbia, Vancouver, British Columbia V6T 1Z1, Canada}

\author{D.~Vavilov}
\affiliation{TRIUMF, 4004 Wesbrook Mall, Vancouver, British Columbia V6T 2A3, Canada}




\collaboration{PIENU Collaboration}

\date{\today}

\begin{abstract}

The three body pion decays ${\pi}^+{\rightarrow}l^+{\nu}X~(l=e,{\mu})$, where $X$ is a weakly interacting neutral boson, were searched for using the full data set  from the PIENU experiment. 
An improved limit on ${\Gamma}({\pi}^+{\to}e^+{\nu}X)/{\Gamma}({\pi}^+{\to}{\mu}^+{\nu}_{\mu})$ in the mass range $0<m_X<120$ MeV/$c^2$ and a first result for ${\Gamma}({\pi}^+{\to}{\mu}^+{\nu}X)/{\Gamma}({\pi}^+{\to}{\mu}^+{\nu}_{\mu})$ in the region $0<m_X<33.9$ MeV/$c^2$ were obtained. 
The Majoron-neutrino coupling model was also constrained using the current experimental result of the ${\pi}^+{\to}e^+{\nu}_e({\gamma})$ branching ratio.

\end{abstract}

\maketitle


\section{\label{sec:Introduction}Introduction}

The existence of massive or massless weakly interacting neutral particles ($X$) has been suggested to augment the standard model with motivations that include providing dark matter candidates \cite{DarkMatter}, explaining baryogenesis \cite{Baryogenesis}, revealing the origin of neutrino masses \cite{SK}, and finding solutions to the strong $CP$ problem \cite{StrongCP1, StrongCP2} involving the axion \cite{familon, axion1, axion2, axion3, axion4}. 
Pion and kaon decays are potential sources of $X$ particles as discussed by Altmannshofer, Gori, and Robinson \cite{ALP} who investigated a model with axionlike particles involved in pion decay ${\pi}^+{\to}e^+{\nu}X$. 
Batell {\it et al.} \cite{DM} studied a model of thermal dark matter emitted in three body meson decay ${\pi}^+(K^+){\to}l^+{\chi}{\phi}$ where $\chi$ and $\phi$ are assumed to be sterile neutrinos. 
Light vector bosons emitted in ${\pi}^+(K^+){\to}l^+{\nu}X$ decay have been discussed by Dror \cite{Dror}.

A Nambu-Goldstone boson, the ``Majoron" proposed by Gelmini and Roncadelli \cite{majoron1}, is also a  candidate of interest. 
It arises in gauge models that have a spontaneous breaking of the baryon and lepton numbers ($B-L$) global symmetry \cite{majoron1, majoron2}. 
In the Majoron models, neutrino masses arise from the vacuum expectation value of a weak isotriplet scalar Higgs boson. 
Barger, Keung, and Pakvasa extended the Majoron model to the decay processes of pions and kaons ${\pi}^+(K^+){\to}l^+{\nu}X$ via  Majoron-neutrino couplings \cite{majoron3}. 
Other related processes and models have been discussed in Refs. \cite{ref1, ref2, ref3, ref4}.

\begin{figure}
\includegraphics[scale=0.45]{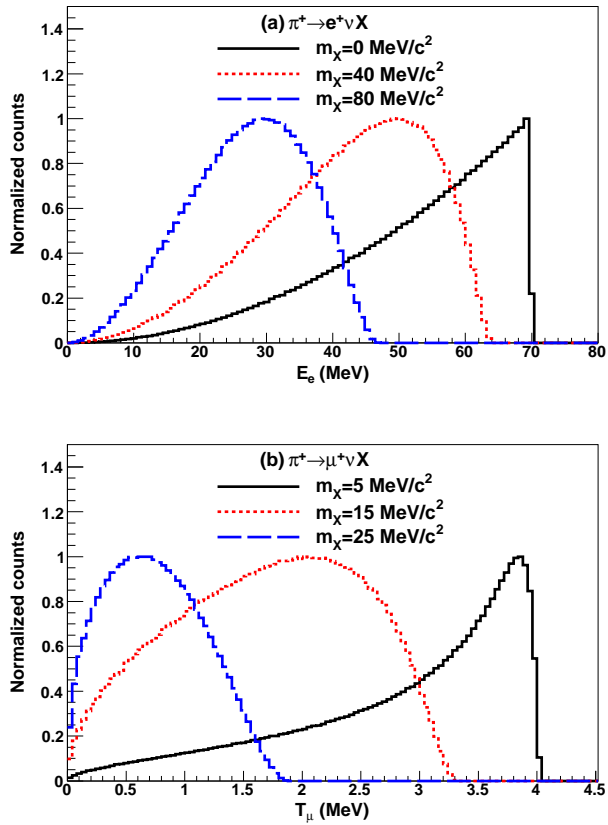}
\caption{\label{fig:MajoronShapes} Total energy spectra of ${\pi}^+{\to}e^+{\nu}X$ and kinetic energy spectra of ${\pi}^+{\to}{\mu}^+{\nu}X$ decays. (a) ${\pi}^+{\to}e^+{\nu}X$ decay with mass $m_X$ of 0 MeV/$c^2$ (solid black), 40 MeV/$c^2$ (dotted red), and 80 MeV/$c^2$ (dashed blue). (b) ${\pi}^+{\to}{\mu}^+{\nu}X$ decay with mass $m_X$ of 5 MeV/$c^2$ (solid black), 15 MeV/$c^2$ (dotted red), and 25 MeV/$c^2$ (dashed blue).}
\end{figure}

Three body pion decays ${\pi}^+{\to}l^+{\nu}X$ can be investigated using the decay lepton energy spectra in pion decays. 
Figure \ref{fig:MajoronShapes} shows the total and kinetic energy spectra of ${\pi}^+{\to}e^+{\nu}X$ and ${\pi}^+{\to}{\mu}^+{\nu}X$ decays assuming the decay products of $X$ are invisible or have very long lifetimes  allowing undetected escape. 
The signal shapes were obtained from Eq. (12) in Ref. \cite{DM}. 
A previous search for the decay ${\pi}^+{\rightarrow}e^+{\nu}X$ was performed by Picciotto {\it et al.} \cite{Picciotto} as a  byproduct of the branching ratio measurement $R^{\pi}={\Gamma}[{\pi}^+{\to}e^+{\nu_e}({\gamma})]/{\Gamma}[{\pi}^+{\rightarrow}{\mu}^+{\nu_{\mu}}({\gamma})]$, where ($\gamma$) indicates the inclusion of radiative decays, using stopped pions in an active target \cite{Britton}. 
The upper limit on the branching ratio was found to be $R^{{\pi}e{\nu}X}={\Gamma}({\pi}^+{\to}e^+{\nu}X)/{\Gamma}({\pi}^+{\to}{\mu}^+{\nu_{\mu}}){\lesssim}4{\times}10^{-6}$ in the mass range  $m_X$ from 0 to 125 MeV/$c^2$. 
The sensitivity was limited by statistics and the remaining background originated from pion decay-in-flight (${\pi}$DIF) events. 
For ${\pi}^+{\to}{\mu}^+{\nu}X$ decay, no comparable studies have been performed. 

In the present work, the decays ${\pi}^+{\to}e^+{\nu}X$ and ${\pi}^+{\to}{\mu}^+{\nu}X$ were sought using the full data set of the PIENU experiment \cite{PIENU} corresponding to two orders of magnitude larger statistics than the previous experiment \cite{Picciotto}. 
The analyses were based on the searches for heavy neutrinos ${\nu}_H$ in ${\pi}^+{\to}e^+{\nu_H}$ decay \cite{PIENU2} and ${\pi}^+{\to}{\mu}^+{\nu_H}$ decay \cite{PIENU3}, and the decays ${\pi}^+{\to}e^+{\nu}_e{\nu}\bar{\nu}$ and ${\pi}^+{\to}{\mu}^+{\nu}_{\mu}{\nu}\bar{\nu}$ \cite{PIENU4}.

\section{Experiment}

\begin{figure}
\includegraphics[scale=0.4, clip, trim=7cm 0cm 7cm 0cm]{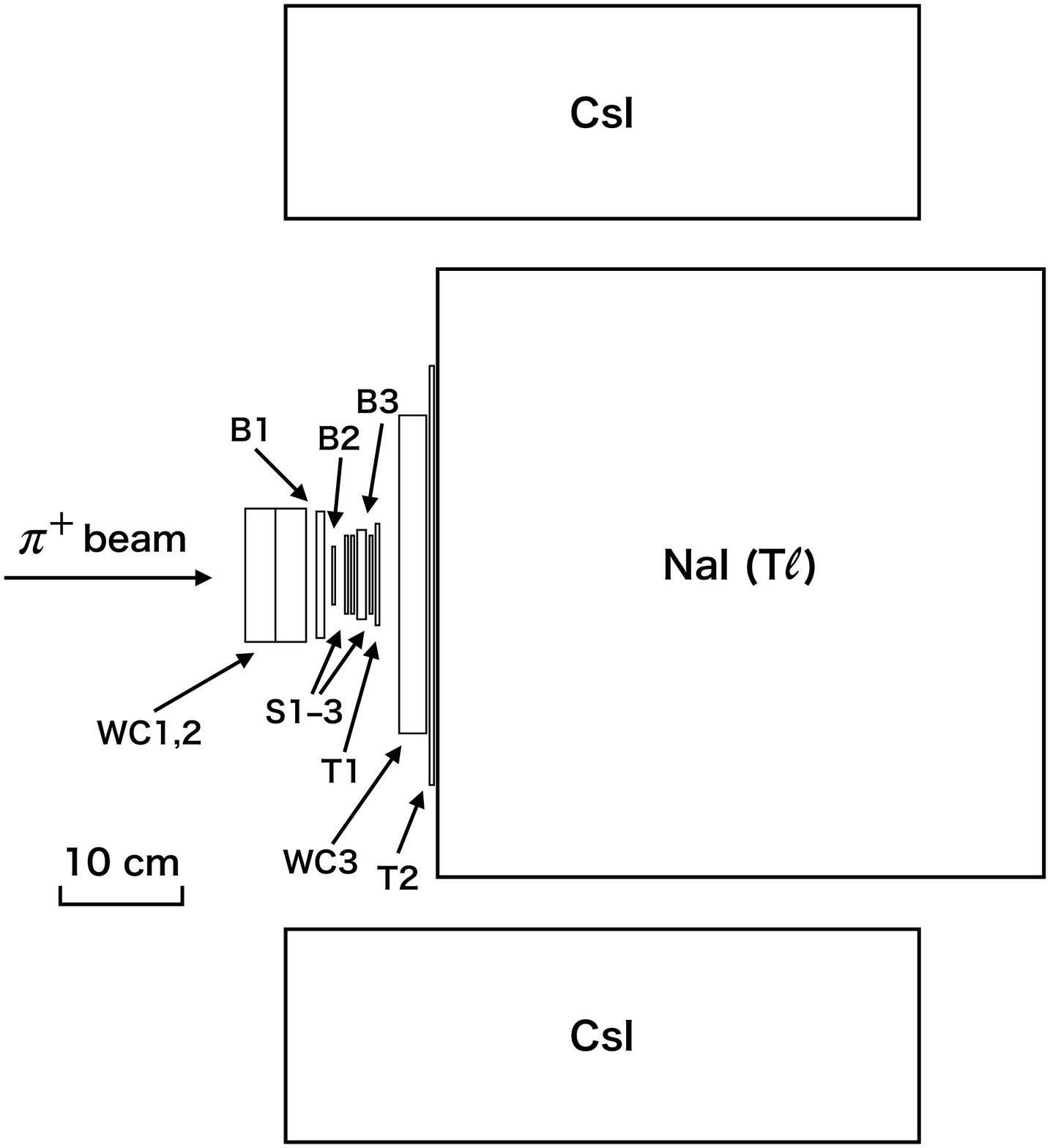}
\caption{\label{fig:detector} Schematic of the PIENU detector \cite{NIMA}.}
\end{figure}

The PIENU detector \cite{NIMA} shown schematically in Fig. \ref{fig:detector} was designed to measure the pion branching ratio $R^{\pi}={\Gamma}[{\pi}^+{\to}e^+{\nu_e}({\gamma})]/{\Gamma}[{\pi}^+{\rightarrow}{\mu}^+{\nu_{\mu}}({\gamma})]$. 
The decay positron in ${\pi}^+{\rightarrow}e^+{\nu_e}$ decay has total energy $E_e=69.8$ MeV. 
For ${\pi}^+{\to}{\mu}^+{\nu_{\mu}}$ decay followed by ${\mu}^+{\to}e^+{\nu_e}\bar{\nu_{\mu}}$ decay (${\pi}^+{\to}{\mu}^+{\rightarrow}e^+$ decay chain), the decay muon has kinetic energy $T_{\mu}=4.1$ MeV and a range in plastic scintillator of about 1 mm; the total energy of the positron in the subsequent muon decay ${\mu}^+{\to}e^+{\nu}_e\bar{\nu}_{\mu}$ ranges from $E_e=0.5$ to 52.8 MeV. 

A pion beam with momentum of $75{\pm}1$ MeV/$c$ provided by the TRIUMF M13 beam line \cite{M13} was tracked by two multiwire proportional chambers (WC1 and WC2) and two sets of silicon strip detectors (S1 and S2). 
Following WC2, the beam was degraded by two thin plastic scintillators (B1 and B2) to measure time and energy loss for particle identification. 
After S2, pions stopped and decayed at rest in the center of an 8 mm thick plastic scintillator target (B3). 
The pion stopping rate in B3 was $5{\times}10^4$ ${\pi}^+/$s. 

Positrons from pion or muon decay were detected by another silicon strip detector (S3) and a multiwire proportional chamber (WC3) located downstream of B3 to reconstruct tracks and define the acceptance. 
Two thin plastic scintillators (T1 and T2) were used to measure the positron time, and its energy was measured by a 48 cm (dia.) $\times$ 48 cm (length) single crystal NaI(T$\ell$) calorimeter surrounded by 97 pure CsI crystals to detect shower leakage. 
The energy resolution of the calorimeter for positrons was 2.2\% (FWHM) at 70 MeV. 

The pion and positron signals were defined by a coincidence of B1, B2, and B3, and a coincidence of T1 and T2, respectively. 
A coincidence of the pion and positron signals within a time window of $-$300 ns to 540 ns with respect to the pion signal was the basis of the main trigger condition. 
This was prescaled by a factor of 16 to form an unbiased trigger (Prescaled trigger). 
 ${\pi}^+{\rightarrow}e^+{\nu_e}$ event collection was enhanced by an early time trigger selecting all events occurring between 6 and 46 ns after the arrival of the pion (Early trigger).  
The typical trigger rate including calibration triggers was about 600 s$^{-1}$. 

To extract the energy and time information, plastic scintillators, silicon strip detectors and CsI crystals, and the NaI(T$\ell$) crystal were read out by 500 MHz, 60 MHz, and 30 MHz digitizers, respectively.
The wire chambers and trigger signals were read by multi-hit time$-$to$-$digital converters with 0.625 ns resolution \cite{NIMA}.

\section{\label{sec:pie} ${\pi}^+{\rightarrow}e^+{\nu}X$ decay}

\subsection{\label{subsec:pieselection} Event selection}

\begin{figure*}
\begin{minipage}[b]{1\linewidth}
\includegraphics[scale=0.68]{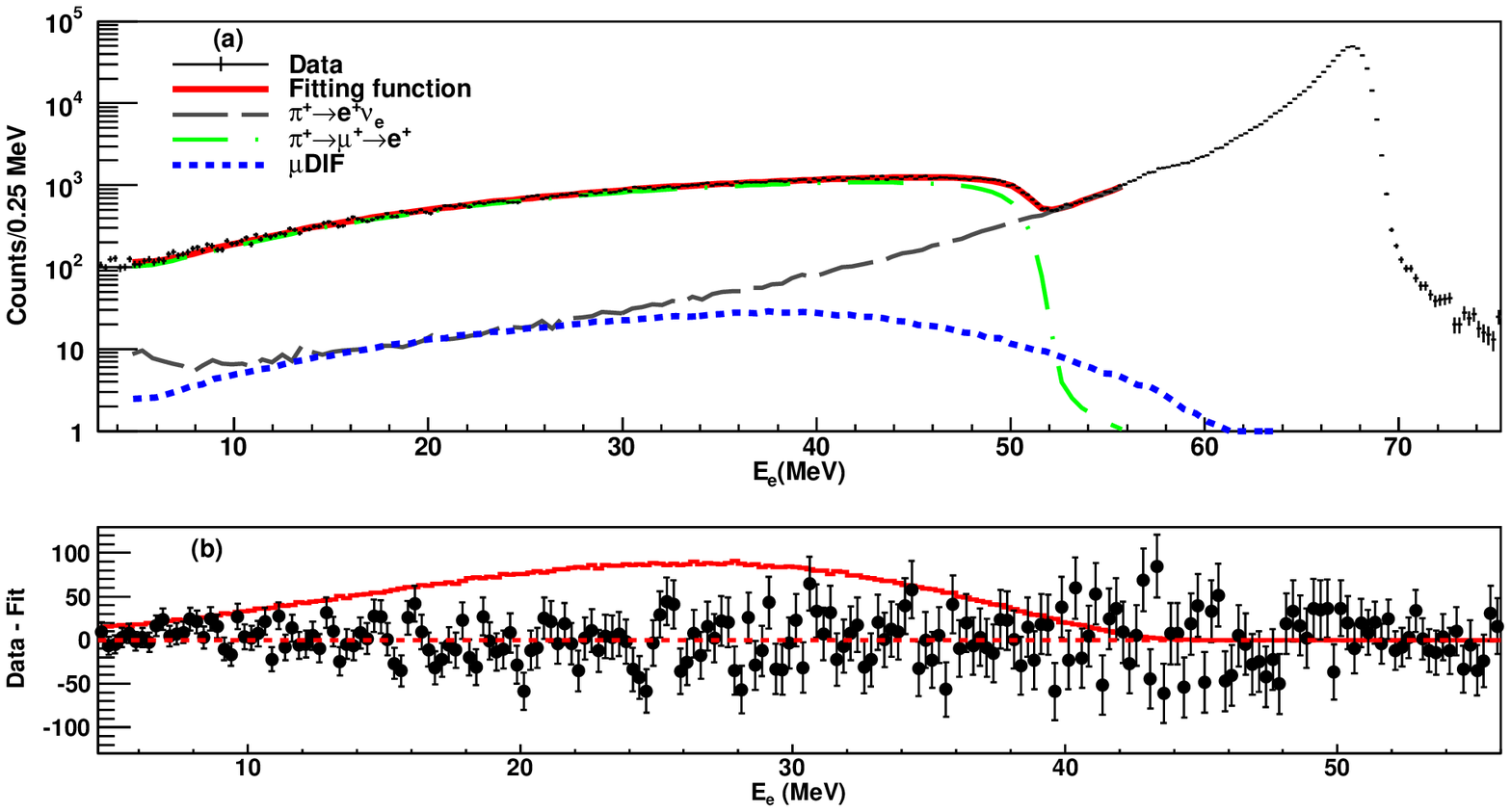}
\end{minipage}\\
\begin{minipage}[b]{1\linewidth}
\includegraphics[scale=0.68]{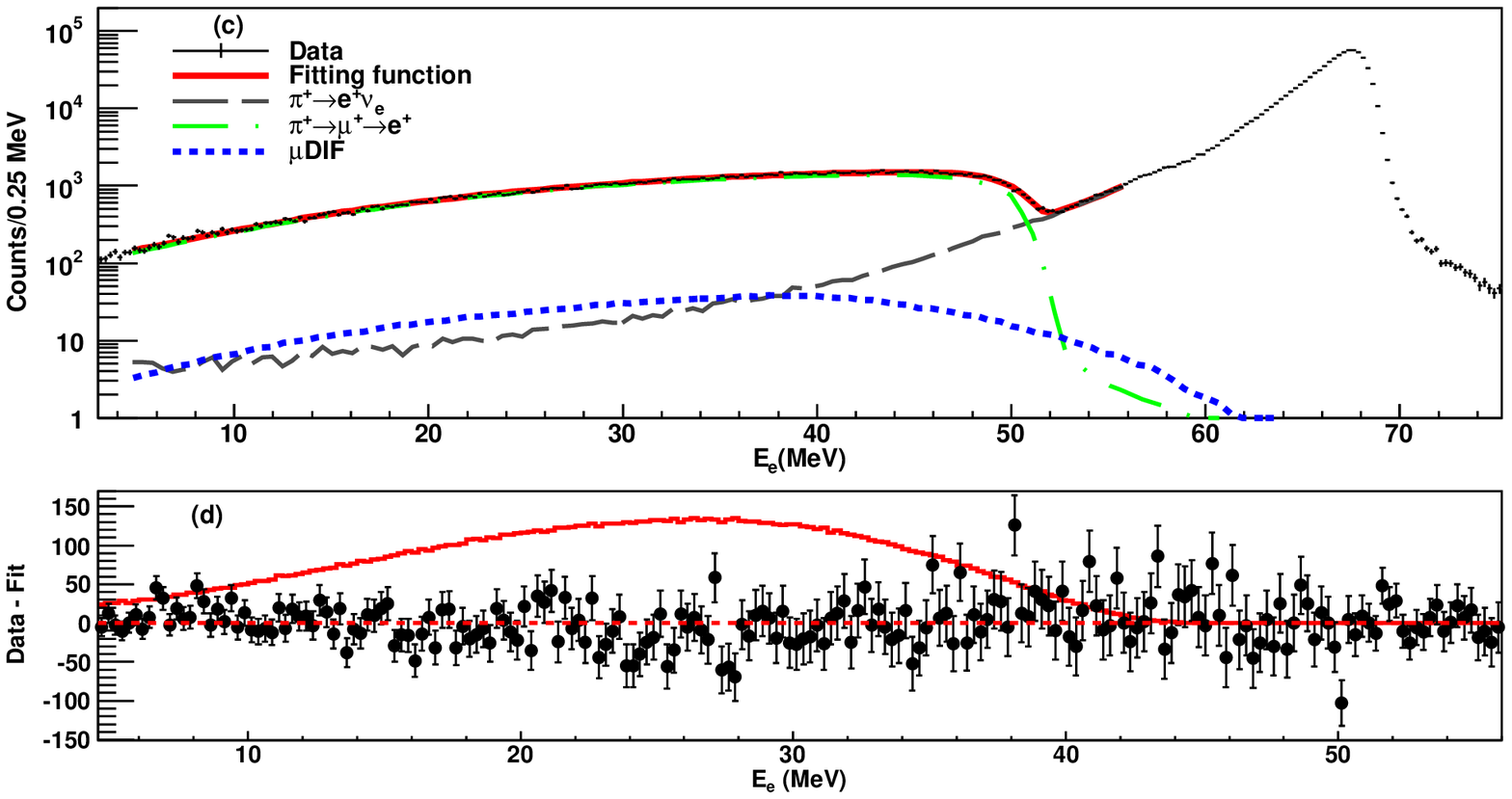}
\end{minipage}
\caption{\label{fig:Suppressed} First and third panels from the top: the $E_e$ spectra of ${\pi}^+{\to}e^+{\nu_e}$ decay after ${\pi}^+{\rightarrow}{\mu}^+{\rightarrow}e^+$ suppression cuts for datasets 1 (a) and 2 (c). The black crosses with the statistical uncertainties show the data. Background components illustrated by the dashed and dotted green line, dotted blue line, dashed gray line, and solid red line represent ${\pi}^+{\rightarrow}{\mu}^+{\rightarrow}e^+$ decays, low energy ${\pi}^+{\rightarrow}e^+{\nu_e}$ tail, $\mu$DIF events, and the sum of those three components, respectively (see text). Second and fourth panels from the top: the residual plots shown by the black circles with statistical error bars and hypothetical signals (solid red lines) with a mass of $m_X=80$ MeV/$c^2$ and a branching ratio $R^{{\pi}e{\nu}X}=2.0{\times}10^{-6}$ from datasets 1 (b) and 2 (d) (the branching ratio obtained by the fit at this mass was $R^{{\pi}e{\nu}X}=(-7.1{\pm}7.1){\times}10^{-8})$.}
\end{figure*}

The decay ${\pi}^+{\rightarrow}e^+{\nu}X$ was searched for by fitting the ${\pi}^+{\rightarrow}e^+{\nu_e}$ energy spectra after ${\pi}^+{\rightarrow}{\mu}^+{\rightarrow}e^+$ background suppression. 
The cuts used for the pion selection, the rejection of the extra activity in scintillators, and the suppression of ${\pi}^+{\to}{\mu}^+{\to}e^+$ backgrounds were the same as for the analysis of ${\pi}^+{\to}e^+{\nu}_e{\nu}\bar{\nu}$ decay \cite{PIENU4}. 
Pions were identified using the energy loss information in B1 and B2. 
Events with extra activity in B1, B2, T1 or T2 were rejected. 
Since the calibration system for the CsI crystals was not available before November 1, 2010, the data were divided into two sets (dataset 1, before, and dataset 2, after November 1, 2010). 
A 15\%  solid angle cut was used for the dataset 2, and a tighter cut (10\%) was applied to the dataset 1 to minimize the effects of electromagnetic shower leakage. 

The ${\pi}^+{\rightarrow}{\mu}^+{\rightarrow}e^+$ backgrounds were suppressed using decay time, energy in the target, and tracking information provided by WC1, WC2, S1, and S2 \cite{PIENU3, PIENU4}. 
Events were first selected by the Early trigger and a decay time cut $t=7-35$ ns after the pion stop was applied. 
The energy loss information in B3 was used because ${\pi}^+{\rightarrow}{\mu}^+{\rightarrow}e^+$ backgrounds deposit larger energy in B3 than ${\pi}^+{\to}e^+{\nu}_e$ decays due to the presence of the decay muon ($T_{\mu}=4.1$ MeV). 
After the timing selection and the energy cut in B3, the beam pion tracking cut, which used the angle between WC1, 2 and S1, 2 track segments, was applied to reject events with a larger angle than most  ${\pi}^+{\to}e^+{\nu}_e$ events (mostly, $\pi$DIF events before B3) \cite{NIMA}. 
Figure \ref{fig:Suppressed} shows the decay positron energy spectra of ${\pi}^+{\rightarrow}e^+{\nu_e}$ decays after ${\pi}^+{\rightarrow}{\mu}^+{\rightarrow}e^+$ background suppression cuts ((a)  dataset 1 and (c) dataset 2). 
The bumps in the positron energy spectra at about 58 MeV are due to photo-nuclear reactions in the NaI(T$\ell$) \cite{PN}. 
The total number of ${\pi}^+{\rightarrow}e^+{\nu_e}$ events was $1.3{\times}10^6$ ($5{\times}10^{5}$ in dataset 1 and $8{\times}10^{5}$ in dataset 2).

\subsection{Energy spectrum fit}

The energy spectrum was fitted with a combination of background terms and a shape to represent the signal.
The background component due to the remaining ${\pi}^+{\rightarrow}{\mu}^+{\rightarrow}e^+$ events was obtained from the data by requiring a late time region $t>200$ ns. 
The shape of the low energy ${\pi}^+{\rightarrow}e^+{\nu_e}$ tail was obtained by Monte Carlo (MC)  simulation \cite{geant4} including the detector response which was measured using a mono-energetic positron beam \cite{NIMA, PN}. 
Because the solid angle cut was reduced and the CsI was not used for dataset 1, the  shapes of the low energy ${\pi}^+{\to}e^+{\nu}_e$ tails are slightly different for the two datasets. 
Another background came from the decays-in-flight of muons ($\mu$DIF) following ${\pi}^+{\rightarrow}{\mu}^+{\nu_{\mu}}$ decays in B3 that has a similar time distribution to ${\pi}^+{\rightarrow}e^+{\nu_e}$ decay. 
The shape of the $\mu$DIF event spectrum was obtained by MC simulation. 
The signal shapes as shown in Fig. \ref{fig:MajoronShapes} (a) were produced with mass range $m_X$ from 0 to 120 MeV/$c^2$ in 5 MeV/$c^2$ steps by MC simulation including the detector response. 
These shapes were normalized to 1 and used for the fit to search for the signals. 
To combine the two data sets, simultaneous fitting with a common branching ratio as a free parameter was performed. 
The fit in the range of $E_e=5-56$ MeV without any signal resulted in ${\chi}^2/$d.o.f.=1.04 (d.o.f.=402). 
The addition of the signals did not change the fit result.

\subsection{Results}

\begin{figure}
\includegraphics[scale=0.45]{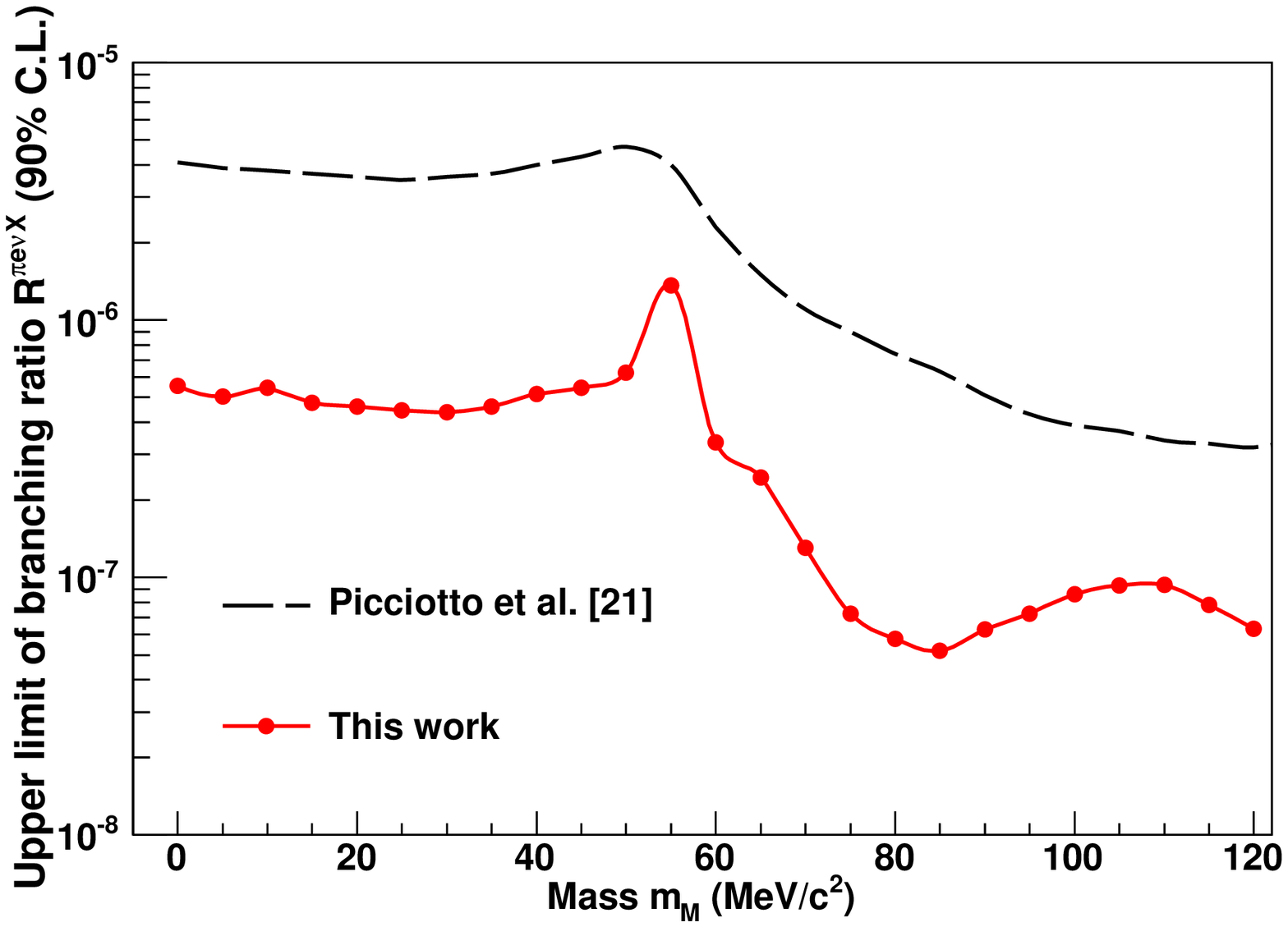}
\caption{\label{fig:BR_UL} Results of the 90\% C.L. upper limit branching ratio $R^{{\pi}e{\nu}X}$. Dashed black line: previous TRIUMF results \cite{Picciotto}. Solid red line with filled circles: results from this work.}
\end{figure}

Figure \ref{fig:Suppressed} (b) and (d) show the residual plots without any signal in datasets 1 and 2;  hypothetical signals assuming $m_X=80$ MeV/$c^2$ with the branching ratio $R^{{\pi}e{\nu}X}=2.0{\times}10^{-6}$ are also shown. 
No significant excess above the statistical uncertainty was observed. 
For example, the branching ratio with $m_X=0$ MeV/$c^2$ obtained by the fit was $R^{{\pi}e{\nu}X}=(0.3{\pm}3.2){\times}10^{-7}$. 
Figure \ref{fig:BR_UL} shows the 90\% confidence level (C.L.) upper limits for the branching ratio ${\pi}^+{\rightarrow}e^+{\nu}X$ in the mass region from 0 to 120 MeV/$c^2$ calculated using the Feldman and Cousins (FC) approach \cite{FC}. 
Since the signal shape at a mass of 55 MeV/$c^2$ is similar to the ${\pi}^+{\to}{\mu}^+{\to}e^+$ energy spectrum, the sensitivity was worse than for other masses due to the strong correlation;  $R^{{\pi}e{\nu}X}=(-0.3{\pm}10.0){\times}10^{-7}$. 
The statistical uncertainty dominates because the systematic uncertainties and the acceptance effects are approximately canceled out by taking the ratio of the number of signal events obtained by the fit to the number of pion decays. 
The acceptance effect due to the cuts was examined by generating positrons in B3 isotropically with an energy range of $E_e=0-70$ MeV using the MC simulation and the systematic uncertainty was estimated to be $<$5\%. 
Compared to the previous TRIUMF experiment \cite{Picciotto}, the limits were improved by an order of magnitude.

\section{\label{muon}${\pi}^+{\rightarrow}{\mu}^+{\nu}X$ decay}

The decay ${\pi}^+{\rightarrow}{\mu}^+{\nu}X$ can be sought by a measurement of the muon kinetic energy in ${\pi}^+{\to}{\mu}^+{\nu}$ decay (followed by ${\mu}^+{\to}e^+{\nu}_e\bar{\nu}_{\mu}$ decay) in the target (B3). 
In the ${\pi}^+{\to}{\mu}^+{\to}e^+$ decay chain, three hits are expected in B3: the first signal is from the beam pion, the second is from the decay muon, and the third is from the decay positron. 
Thus, the second of three pulses in B3 would be due to the muon kinetic energy. 
However, the pulse detection logic could not efficiently identify  pulses below 1.2 MeV \cite{PIENU2}. 
Therefore, the search was divided into two muon energy regions, above and below 1.2 MeV. 
The number of Prescaled trigger events used for the analysis was $4{\times}10^9$.  
The analysis strategy and event selection cuts were based on the massive neutrino \cite{PIENU3} and three neutrino decay \cite{PIENU4} searches, briefly described in the following sections.

\subsection{\label{sec:MuH} Analysis of the region above 1.2 MeV}

\begin{figure}
\includegraphics[scale=0.45]{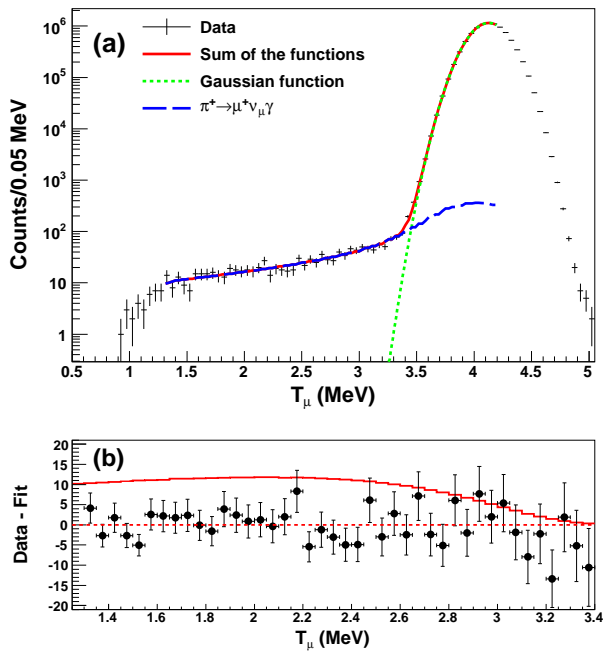}
\caption{\label{fig:FitH} (a) The $T_{\mu}$ spectra of ${\pi}^+{\to}{\mu}^+{\to}e^+$ decay. The black crosses with the statistical uncertainties show the data. The dotted green line, dashed blue line, and solid red line represent a Gaussian distribution centered at 4.1 MeV, ${\pi}^+{\rightarrow}{\mu}^+{\nu}_{\mu}{\gamma}$ decay, and the sum of those two functions, respectively. (b) Residual plots shown by the black circles with statistical error bars in the range $T_{\mu}$=1.3 to 3.4 MeV. The solid red line represents a hypothetical signal with mass of $m_X=15$ MeV/$c^2$ and the branching ratio $R^{{\pi}{\mu}{\nu}X}=6.0{\times}10^{-5}$; the branching ratio obtained by the fit was $R^{{\pi}{\mu}{\nu}X}=(-3.6{\pm}5.1){\times}10^{-6}$.}
\end{figure}

As described in Sec. \ref{subsec:pieselection}, pions were identified using B1 and B2 and events with extra hits in B1, B2, T1, or T2 were rejected. 
A solid angle acceptance of about 20\% for the decay positron was used. 
To ensure the selected events were from ${\pi}^+{\rightarrow}{\mu}^+{\rightarrow}e^+$ decays, a late positron decay time $t>200$ ns after the pion stop  and the positron energy in the NaI(T$\ell$) calorimeter $E_e<55$ MeV were required. 
Then, the events with three clearly separated pulses in the target (B3) were selected and the second pulse information was extracted and assigned to the decay muon \cite{PIENU2}.
The muon kinetic energy ($T_{\mu}$) spectrum after the event selection cuts is shown in Fig. \ref{fig:FitH} (a). 
As described above, the drop below 1.2 MeV was due to the inefficiency of the pulse detection logic \cite{PIENU2}. 
The main background below 3.4 MeV was due to the radiative pion decay ${\pi}^+{\rightarrow}{\mu}^+{\nu_{\mu}}{\gamma}$ (branching fraction $2{\times}10^{-4}$ \cite{pimunug}). 
The total number of ${\pi}^+{\to}{\mu}^+{\to}e^+$ events available was 9.1${\times}10^6$. 

The decay ${\pi}^+{\rightarrow}{\mu}^+{\nu}X$ was searched for by fitting the $T_{\mu}$ energy spectrum of ${\pi}^+{\to}{\mu}^+{\to}e^+$ decays. 
The fit was performed using a Gaussian peak centered at 4.1 MeV (energy resolution ${\sigma}=0.16$ MeV), the ${\pi}^+{\rightarrow}{\mu}^+{\nu}_{\mu}{\gamma}$ decay spectrum obtained by MC simulation \cite{geant4}, and the normalized signal spectra including the energy resolution in B3. 
The signal spectra as shown in Fig. \ref{fig:MajoronShapes} (b) were generated with the mass range  $0<m_X<26$ MeV/$c^2$ with 1 MeV/$c^2$ steps using MC including detector resolution. 
The fit for $T_{\mu}$ from 1.3 to 4.2 MeV without any ${\pi}^+{\rightarrow}{\mu}^+{\nu}X$ signal introduced gave  ${\chi}^2/$d.o.f.=1.27 (d.o.f.=53) and the residuals of the fit for the signal sensitive region are shown in Fig. \ref{fig:FitH} (b). 
The addition of signal components did not change the fit result. 

\begin{figure}
\includegraphics[scale=0.45]{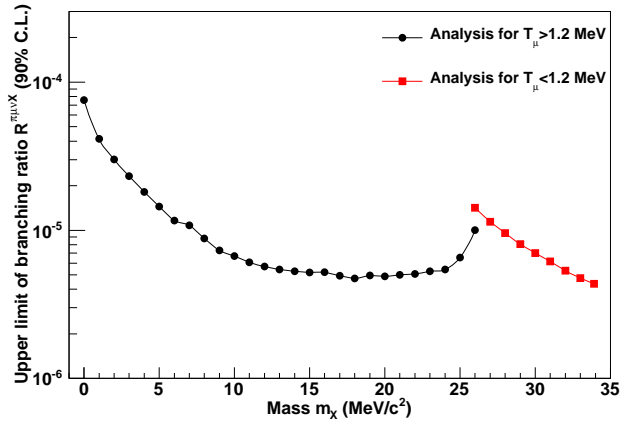}
\caption{\label{fig:BR_UL_Mu_FC} Summary of the 90\% C.L. upper limit branching ratio $R^{{\pi}{\mu}{\nu}X}$ in this work. The black circles show the result of the search in the energy region $T_{\mu}>1.2$ MeV (see text in Sec. \ref{sec:MuH}) and the red squares represent the analysis result in the region $T_{\mu}<1.2$ MeV (see text in Sec. \ref{sec:MuL}).}
\end{figure}

No significant signal beyond the statistical uncertainty was observed. 
For example, the branching ratios for the signals with mass $m_X=0$ MeV/$c^2$ and 26 MeV/$c^2$ obtained by the fit were $R^{{\pi}{\mu}{\nu}X}={\Gamma}({\pi}^+{\to}{\mu}^+{\nu}X)/{\Gamma}({\pi}^+{\to}{\mu}^+{\nu}_{\mu})=(-2.1{\pm}1.3){\times}10^{-4}$ and $(-4.8{\pm}8.8){\times}10^{-6}$, respectively. 
Systematic uncertainties and acceptance effects were approximately canceled by taking the ratio of amplitudes for the signal and ${\pi}^+{\to}{\mu}^+{\nu}_{\mu}$ decays. 
The systematic uncertainties and acceptance effects due to the cuts were examined by generating decay muons in the target with several kinetic energies in the range $T_{\mu}=0-4.1$ MeV using MC simulation, and the systematic uncertainty was estimated to be $<$5\%. 
The black circles in Fig. \ref{fig:BR_UL_Mu_FC} show the result of the 90\% C.L. upper limit branching ratio $R^{{\pi}{\mu}{\nu}X}$ in this energy region calculated using the FC method.

\subsection{\label{sec:MuL} Analysis of the region below 1.2 MeV}

\begin{figure}
\includegraphics[scale=0.45]{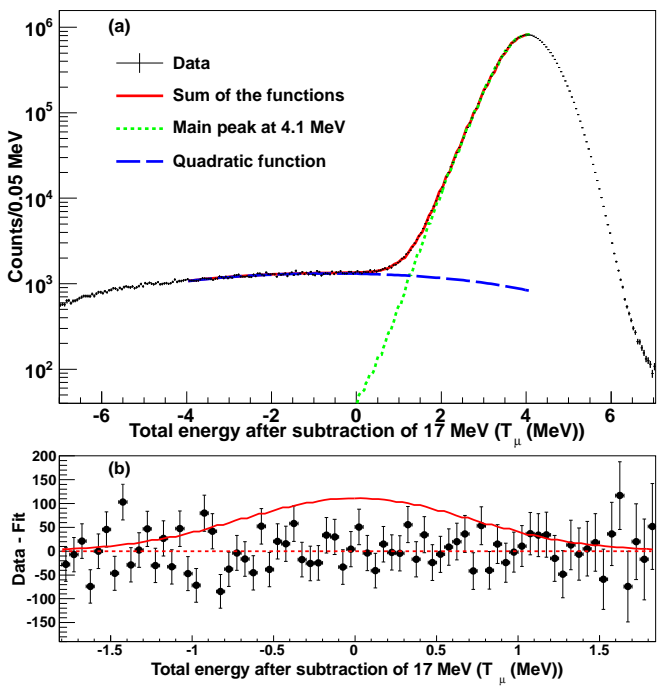}
\caption{\label{fig:FitL} (a) The total energy in the target due to the pion and muon after subtracting 17 MeV. The black crosses with statistical uncertainties show the data. The dotted green line, dashed blue line, and solid red line represent the main peak at 4.1 MeV, quadratic background due to ${\pi}$DIF events, and the sum of those two functions, respectively. (b) Residual plots shown by the black circles with the statistical error bars in the signal region  $T_{\mu}$=-1.8 to 1.8 MeV. The solid red line represents a hypothetical signal with mass of $m_X=33.9$ MeV/$c^2$ and the branching ratio $R^{{\pi}{\mu}{\nu}X}=3.0{\times}10^{-5}$.}
\end{figure}

For $T_{\mu}<1.2$ MeV, the selection of pions, rejection of extra activity in scintillators, the solid angle cut for the decay positron, and the positron energy cut in the NaI(T$\ell$) calorimeter were all the same as in the analysis in the energy region $T_{\mu}>1.2$ MeV. 
To minimize ${\pi}$DIF events, the same tracking cut by WC1, WC2, S1, and S2 used in Sec. \ref{subsec:pieselection} was also applied. 
After these basic cuts, the energies observed in B3 in a wide time window (700 ns) including pion and positron energies were obtained. 
To cleanly subtract the positron contribution from the integrated energy, events with late positron decay $t>300$ ns were selected and the isolated positron energy was subtracted. 
After that, the contribution of the averaged pion kinetic energy ($\sim$17 MeV) was subtracted from the total energy (due to the pion and the muon). 
Figure \ref{fig:FitL} (a) shows the total energy (corresponding to $T_{\mu}$) after subtracting 17 MeV. 
The background below $T_{\mu}<1$ MeV was mainly due to remaining ${\pi}$DIF events. 
The number of ${\pi}^+{\to}{\mu}^+{\to}e^+$ events available for the analysis is $1.3{\times}10^8$.

There are two background shapes, the 4.1 MeV peak and the ${\pi}$DIF events. 
A quadratic function was used for the ${\pi}$DIF events. 
To search for ${\pi}^+{\to}{\mu}^+{\nu}X$ decay, the width of the signal shape was scaled using that at the 4.1 MeV peak. 
Figure \ref{fig:FitL} (b) shows the residual plots in the signal region from -1.8 to 1.8 MeV without any signal shape and a hypothetical signal shape assuming a mass of $m_X=33.9$ MeV/$c^2$ with the branching ratio $R^{{\pi}{\mu}{\nu}X}=3.0{\times}10^{-5}$. 
The branching ratio obtained by the fit was $(1.0{\pm}2.0){\times}10^{-6}$. 
The fit was performed from -4.0 to 4.1 MeV and the fitting range of -4.0 to 2.0 MeV (signal region) resulted in ${\chi}^2/$d.o.f.=1.03 (d.o.f.=115); there is some small deviation above 2 MeV due to a small mismatch due to the kinetic energy distribution of the beam pion. 

The signals of ${\pi}^+{\to}{\mu}^+{\nu}X$ decay were searched for in the mass range of $m_X=26$ to 33.9 MeV/$c^2$, but no significant excess beyond the statistical uncertainty was observed. 
The red squares in Fig. \ref{fig:BR_UL_Mu_FC} represent the result of the 90\% C.L. upper limit branching ratio $R^{{\pi}{\mu}{\nu}X}$ in this energy region calculated using the FC approach.

\section{Constraints on the Majoron model}

The Majoron model can be constrained using the experimental value of the pion branching ratio $R^{\pi}$. 
The predicted branching ratio including the massless Majoron $X_0$ and a light neutral Higgs $H'$ ($\lesssim$1 MeV/$c^2$) can be written as
\begin{equation}
\frac{{\Gamma}({\pi}{\to}eL^0)/{\Gamma}({\pi}{\to}{\mu}L^0)}{{\Gamma}({\pi}{\to}e{\nu}_e)/{\Gamma}({\pi}{\to}{\mu}{\nu}_{\mu})}=1+157.5g^2
\end{equation}
where $L^0$ is the final state ${\nu}$, ${\nu}X_0$, and ${\nu}H'$, and $g$ is the Majoron-neutrino coupling constant \cite{majoron3}. 
The upper limit of the ratio $R^{\pi}_{\rm exp}/R^{\pi}_{\rm SM}$ at 90\% C.L. using the current averaged experimental value $R^{\pi}_{\rm exp}=(1.2327{\pm}0.0023){\times}10^{-4}$ \cite{PDG} is
\begin{equation}
\frac{R^{\pi}_{\rm exp}}{R^{\pi}_{\rm SM}}<1.0014.
\end{equation}
Using this limit, the 90\% C.L. upper limit of the coupling constant can be found to be
\begin{equation}
g^2<9{\times}10^{-6},
\end{equation}
which was improved by a factor of three over the previous experiment \cite{Britton}.

\section{Conclusion}

No evidence of the three body pion decays ${\pi}^+{\to}e^+{\nu}X$ or ${\pi}^+{\to}{\mu}^+{\nu}X$ was found and new upper limits were set. 
The limits on the branching ratio ${\pi}^+{\to}e^+{\nu}X$ were improved by an order of magnitude over the previous experiment. 
For ${\pi}^+{\to}{\mu}^+{\nu}X$ decay, the limits obtained are the first available results. 
The Majoron model was also constrained using the pion branching ratio $R^{\pi}$.

\begin{acknowledgments}
This work was supported by the Natural Sciences and Engineering Research Council of Canada (NSERC, No. SAPPJ-2017-00033), and by the Research Fund for the Doctoral Program of Higher Education of China, by CONACYT doctoral fellowship from Mexico, and by JSPS KAKENHI Grant No. 18540274, No. 21340059, No. 24224006, and No. 19K03888 in Japan.
We are grateful to Brookhaven National Laboratory for the loan of the crystals, and to the TRIUMF operations, detector, electronics and DAQ groups for their engineering and technical support.
\end{acknowledgments}


\begin{thebibliography}{99}

\bibitem{DarkMatter} G. Bertone, D. Hooper, and J. Silk, Phys. Rep. {\bf 405}, 279 (2005).

\bibitem{Baryogenesis} A.D.~Dolgov, arXiv:hep-ph/9707419; V.A.~Rubakov and M.E.~Shaposhnikov, Phys. Usp. {\bf 39}, 461 (1996).

\bibitem{SK} Y. Fukuda {\it et al}., Phys. Rev. Lett. {\bf 81}, (1998) 1562.

\bibitem{StrongCP1} R.D. Peccei and H.R. Quinn, Phys. Rev. Lett. {\bf 38} (1977) 1440.

\bibitem{StrongCP2} R.D. Peccei and H.R. Quinn, Phys. Rev. D {\bf 16} (1977) 1791.

\bibitem{familon} F. Wilczek, Phys. Rev. Lett. {\bf 49}, 1549 (1982); see also A.~Davidson and K. C. Wali, Phys. Rev. Lett. 48, 11 (1982).

\bibitem{axion1} J. Jaeckel and A. Ringwald, Annu. Rev. Nucl. Part. Sci. {\bf 60}, 405 (2010).

\bibitem{axion2} P. Agrawal and K. Howe, J. High Energy Phys. {\bf 12} (2018) 029.

\bibitem{axion3} D.S. M. Alves and N. Weiner, J. High Energy Phys. {\bf 07} (2018) 092.

\bibitem{axion4} K.S. Jeong, T.H. Jung, and C.S. Shin, Phys. Rev. D {\bf 101}, 035009 (2020).


\bibitem{ALP} W. Altmannshofer, S. Gori, and D.J. Robinson, Phys. Rev. D {\bf 101}, 075002 (2020).  

\bibitem{DM} B. Batell, T. Han, D. McKeen, and B.S.E. Haghi, Phys. Rev. D {\bf 97}, 075016 (2018). 

\bibitem{Dror} J.A. Dror, Phys. Rev. D {\bf 101} 095013 (2020).

\bibitem{majoron1} G.B. Gelmini and M. Roncadelli, Phys. Lett. B {\bf 99}, 411 (1981); see also G.B.~Gelmini, S.~Nussinov, and M.~Roncadelli, Nucl. Phys. {\bf B209} (1982) 157-173.

\bibitem{majoron2} Y. Chikashige, R. N. Mohapatra, and R. D. Peccei, Phys. Lett. {\bf 98B}, 265 (1981).

\bibitem{majoron3} V. Barger, W.Y. Keung, and S. Pakvasa, Phys. Rev. D {\bf 25}, 907 (1982).

\bibitem{ref1} A. Masiero, J.W.F. Valle, Phys. Lett. {\bf B251}, 273-278 (1990).

\bibitem{ref2} A.P. Lessa and O.L.G. Peres, Phys. Rev. D {\bf 75}, 094001 (2007).

\bibitem{ref3} M. Hirsch, A. Vicente, J. Meyer, and W. Porod, Phys. Rev. D {\bf79}, 055023 (2009).

\bibitem{ref4} X. Garcia i Tormo, D. Bryman, A. Czarnecki, and M. Dowling, Phys.  Rev. D {\bf 84}, 113010 (2011).

\bibitem{Picciotto} C.E. Picciotto {\it et al}., Phys. Rev. D {\bf 37}, 1131 (1988).

\bibitem{Britton} D.I. Britton {\it et al}., Phys. Rev. Lett. {\bf 68}, 3000 (1992) and Phys. Rev. D {\bf 49}, 28 (1994).

\bibitem{PIENU} A. Aguilar-Arevalo {\it et al}., Phys. Rev. Lett. {\bf 115}, 071801 (2015).

\bibitem{PIENU2} M. Aoki {\it et al}., Phys. Rev. D {\bf 84}, 052002 (2011) and A. Aguilar-Arevalo {\it et al}., Phys. Rev. D {\bf 97}, 072012 (2018).

\bibitem{PIENU3} A. Aguilar-Arevalo {\it et al}., Phys. Lett. B {\bf 798}, 134980 (2019). 

\bibitem{PIENU4} A. Aguilar-Arevalo {\it et al}., Phys. Rev. D {\bf 102}, 012001 (2020).

\bibitem{NIMA} A. Aguilar-Arevalo {\it et al}., Nucl. Instrum. Methods Phys. Res., Sect. A {\bf 791}, 38 (2015).

\bibitem{M13} A. Aguilar-Arevalo {\it et al}., Nucl. Instrum. Methods Phys. Res., Sect. A {\bf 609}, 102 (2009). 

\bibitem{pimunug} G. Bressi, G.~Carugno, S.~Cerdonio, E.~Conti, A.T.~Meneguzzo, and D.~Zanello, Nucl. Phys. B {\bf 513} (1998) 555.

\bibitem{PN} A. Aguilar-Arevalo {\it et al.}, Nucl. Instrum. Methods Phys. Res., Sect. A {\bf 621}, 188 (2010).

\bibitem{geant4} S. Agostinelli {\it et al}. (GEANT4 Collaboration), Nucl. Instrum. Methods Phys. Res., Sect. A {\bf 506}, 250 (2003); http://geant4.cern.ch.

\bibitem{FC} G.J. Feldman and R.D. Cousins, Phys. Rev. D {\bf 57}, 3873 (1998).

\bibitem{PDG}  M. Tanabashi et al. (Particle Data Group), Phys. Rev. D {\bf 98}, 030001 (2018). 

\end{thebibliography}
\end{document}